\documentclass[preprint,review,12pt]{elsarticle}

\usepackage{amssymb}
\usepackage{amsmath}
\usepackage{color}
\newcommand*{\rttensor}[1]{\overline{\overline{#1}}}

\makeatletter
\def\ps@pprintTitle{%
 \let\@oddhead\@empty
 \let\@evenhead\@empty
 \def\@oddfoot{}%
 \let\@evenfoot\@oddfoot}
\makeatother

\journal{Carbon}

\begin{document}

\begin{frontmatter}



\title{Optical modeling, solver, and design of wafer-scale single-enantiomer carbon nanotube film and reconfigurable chiral photonic device}


\author[inst1]{Jichao Fan}

\affiliation[inst1]{organization={Department of Electrical and Computer Engineering, The University of Utah},
            addressline={50 S Central Campus Drive}, 
            city={Salt Lake City},
            postcode={84112}, 
            state={UT},
            country={USA}}

\author[inst1]{Benjamin Hillam}
\author[inst2]{Cheng Guo}

\affiliation[inst2]{organization={Department of Applied Physics, Stanford University},
            addressline={348 Via Pueblo}, 
            city={Stanford},
            postcode={94305}, 
            state={CA},
            country={USA}}

\author[inst3]{Hiroyuki Fujinami}
\author[inst3]{Shiba Koki}
\author[inst1]{Haoyu Xie}
\author[inst1]{Ruiyang Chen}
\author[inst3]{Kazuhiro Yanagi}
\affiliation[inst3]{organization={Department of Physics, Tokyo Metropolitan University},
            addressline={1-1 Minami-Osawa}, 
            city={Hachioji-shi, Tokyo},
            postcode={192-0397}, 
            country={Japan}}

\author[inst1]{Weilu Gao\corref{cor1}}
\cortext[cor1]{Corresponding author: weilu.gao@utah.edu (Weilu Gao)}

\begin{abstract}
The interaction of circularly polarized light with chiral matter and functional devices enables novel phenomena and applications. Recently, wafer-scale solid-state single-enantiomer carbon nanotube (CNT) films have become feasible and are emerging as a chiral photonic material platform thanks to their quantum-confinement-induced optical properties and facile scalable assembly. However, optical modeling, solver, and device design tools for such materials are non-existent. Here, we prepare wafer-scale single-enantiomer (6,5) and (11,-5) randomly oriented CNT films and create an optical material model based on measured experimental optical spectra. We also implement a highly-parallel graphic-processing-unit accelerated transfer matrix solver for general bi-anisotropic materials and layered structures. Further, we demonstrate reconfigurable chiral photonic devices in a heterostructure with phase change materials through machine learning-enabled efficient gradient-based inverse design and optimization. Our developed full stack of a chiral photonic material and device hardware platform and a corresponding high-performance differential-programming-enabled solver opens the door for future chiral photonic devices and applications based on single-enantiomer CNT films.
\end{abstract}




\end{frontmatter}


\section{Introduction}


Manipulating the interaction of circularly polarized light with chiral materials and their functional photonic devices is crucial for many applications, such as sensing~\cite{LiuEtAl2023NRC,WarningEtAl2021N}, imaging~\cite{ZhanEtAl2021AM,KhaliqEtAl2022AOM}, computing~\cite{DanEtAl2024N}, and quantum technologies~\cite{AielloEtAl2022N}. The advancement of developing such devices is nowadays hindered not only by the limited number of high-performance material platforms but also by efficient solvers and design tools. In current material platforms, small natural chiral molecules generally have weak chiroptical responses because of a large mismatch between molecule size and optical wavelength. Chiral organic supramolecular structures~\cite{AlbanoEtAl2020CR,FurlanEtAl2024NP} and twisted structures of anisotropic materials~\cite{HanEtAl2023AM} manufactured through cost-effective bottom-up self-assembly techniques can produce strong and dynamic chiroptical responses, while individual constituents lack quantum-confinement-induced extraordinary properties and the design of structures is largely empirical. Artificial symmetry-breaking periodic structures can provide an engineerable platform with systematic electromagnetic design tools, such as finite difference time domain analysis, while their top-down manufacturing is sophisticated and challenging to scale up~\cite{WangEtAl2016N} and those design tools are generally computationally intensive and time-consuming~\cite{JiangEtAl2020NRM}. 

\begin{figure}
    \centering
    \includegraphics[width=1.0\linewidth]{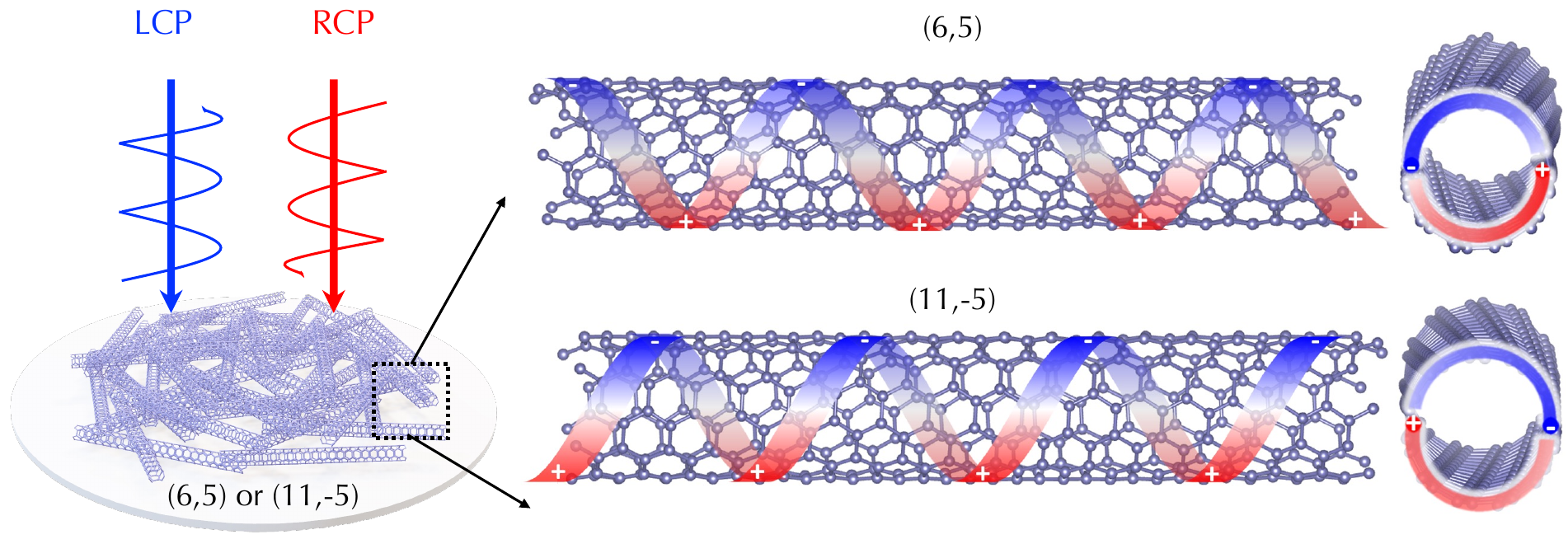}
    \caption{\textbf{Overview}. Illustrations of the interaction of circularly polarized light with single-enantiomer CNT films and their excitonic chiral photonic responses. Schematics were produced using VESTA~\cite{MommaEtAl2011JAC}}
    \label{fig:illustration}
\end{figure}

Carbon nanotubes (CNTs) resurge as a promising chiral photonic material platform in the form of both natural and synthetic chirality thanks to the recent progress in large-scale chirality and enantiomer separation and bottom-up wafer-scale ordered self-assembly techniques~\cite{TanakaEtAl2015AC,WeiEtAl2017JACS,WeiEtAl2018C,LiEtAl2020N,WeiEtAl2022AS,HeEtAl2016NN,GaoEtAl2019RSOS,DoumaniEtAl2023NC}. The optical properties of CNTs are determined by their one-dimensional (1D) atomic structures~\cite{WeismanEtAl2019}. The quantum confinement in 1D nanoscale tubular geometry of CNTs leads to the formation of subbands and the interband transitions between subbands produce strong exciton-dominant optical spectral features. The transition energies can be tuned by the quantum engineering of CNT atomic structures, which are typically described by a pair of chirality indices ($n$, $m$), and by external stimuli such as electric fields and optical excitation. Except for special highly symmetric zigzag ($n$,0) and armchair ($n$, $n$) CNTs, all other CNTs are chiral and contain two enantiomers, denoted as ($n$, $m$) and ($n+m$, $-m$), with different chiral photonic responses. For example, as illustrated in Fig.\,\ref{fig:illustration}, the interaction of (6,5) and (11,-5) CNTs of opposite handednesses with left circularly polarized (LCP) and right circularly polarized (RCP) light, such as circular dichroism (CD) that is the differential absorption of LCP and RCP light, is distinctly different. Hence, their assembled wafer-scale films become a solid-state material platform for building novel chiral photonic devices; see Fig.\,\ref{fig:illustration}. However, to date, there is no demonstration of any optical modeling or solvers for such materials, hindering device design capability.


 

Here, we present an optical material model and a highly-parallel fast solver for wafer-scale single-enantiomer (6,5) and (11,-5) randomly oriented CNT films, and demonstrate reconfigurable chiral photonic devices in a heterostructure with phase change materials (PCMs) through machine learning (ML)-enabled efficient gradient-based inverse design and optimization. Specifically, we prepared single-enantiomer (6,5) and (11,-5) suspensions through gel chromatography and wafer-scale randomly oriented films through vacuum filtration, and experimentally measured CD and absorption spectra of suspension and solid-state samples. We created material models for (6,5) and (11,-5) films and implemented a graphic-processing-unit (GPU) accelerated transfer matrix solver for general bi-anisotropic materials and layered architectures based on a state-of-the-art ML framework. By simultaneously fitting experimental spectra, we extracted dielectric functions and chirality parameters for (6,5) and (11,-5) films. Furthermore, we leveraged the backpropagation algorithm in the ML framework to design a novel reconfigurable chiral photonic device based on a heterostructure consisting of single-enantiomer CNT films and PCM films. Our developed full stack of a chiral photonic material and device hardware platform and a corresponding GPU-accelerated and differential-programming-enabled solver opens the door for future chiral photonic devices and applications based on single-enantiomer CNT films.

\section{Experimental section}

\subsection{Preparation of single-enantiomer CNT suspensions}

We purchased CoMoCAT CNT powders and mixed them with a combination of surfactants. The mixture was tip-sonicated to disperse CNTs to form a homogenized aqueous suspension. The suspension was then purified through ultracentrifugation to remove undispersed impurities, such as catalysts, and large bundles. The supernatant was collected for enantiomer separation, which was performed through an automatic temperature-controlled chromatography system for large-scale separation following the procedures described in Ref.\,\cite{WeiEtAl2017JACS,WeiEtAl2018C,YomogidaEtAl2016NC}. By controlling the composition and concentration of surfactant elution solutions, CNTs of different diameters and handednesses were collected at various times during the process; see more details in \emph{Supplementary Information Note 1}. Fig.\,\ref{fig:exp}a displays a photograph of one prepared single-enantiomer suspension. By repeating the enantiomer separation process multiple times, we collected enough CNT suspensions for preparing wafer-scale films.

\subsection{Preparation of single-enantiomer CNT films}

We produced wafer-scale (e.g., $5\times5$\,mm$^2$) single-enantiomer (6,5) and (11,-5) CNT films using vacuum filtration. Briefly, we exchanged the mixed surfactant system in separated single-enantiomer suspensions to a single sodium deoxycholate (DOC) surfactant system before pouring it into the filtration system. The DOC concentration was kept higher than its critical micelle concentration and a fast filtration was performed to produce randomly oriented films instead of aligned ones~\cite{HeEtAl2016NN}. When all suspensions finished the filtration process, the deposited films were kept on the filtration system for a complete dry. The obtained films on the filter membrane can be transferred onto arbitrary substrates following a wet transfer process~\cite{HeEtAl2016NN}; see more details in \emph{Supplementary Information Note 2}. For optical characterization, we transferred films onto silica substrates, which have broadband optical transparency from ultraviolet to near-infrared wavelength ranges. Fig.\,\ref{fig:exp}b shows a photo of one produced film. Due to the self-limiting nature of the vacuum filtration process~\cite{WuEtAl2004S,ZhaoEtAl2024S}, the film thickness was uniform and can be controlled by adjusting the suspension concentration and volume. 

\subsection{Film thickness measurement and optical characterization of anisotropy}

We characterized the thicknesses of obtained films through atomic force microscopy (Bruker Multimode 8 AFM). Supplementary Fig.\,1a displays a representative AFM image of one CNT film. The film thickness was measured by creating a section profile that crossed a crack in the film, leading to a relatively smooth edge. In contrast, multiple bumps formed around the physical edge of the film because the film was cut into certain shapes manually. Supplementary Fig.\,1b shows the height profile of the film with a thickness $\sim60\,$nm. 

\begin{figure}
    \centering
    \includegraphics[width=1.0\linewidth]{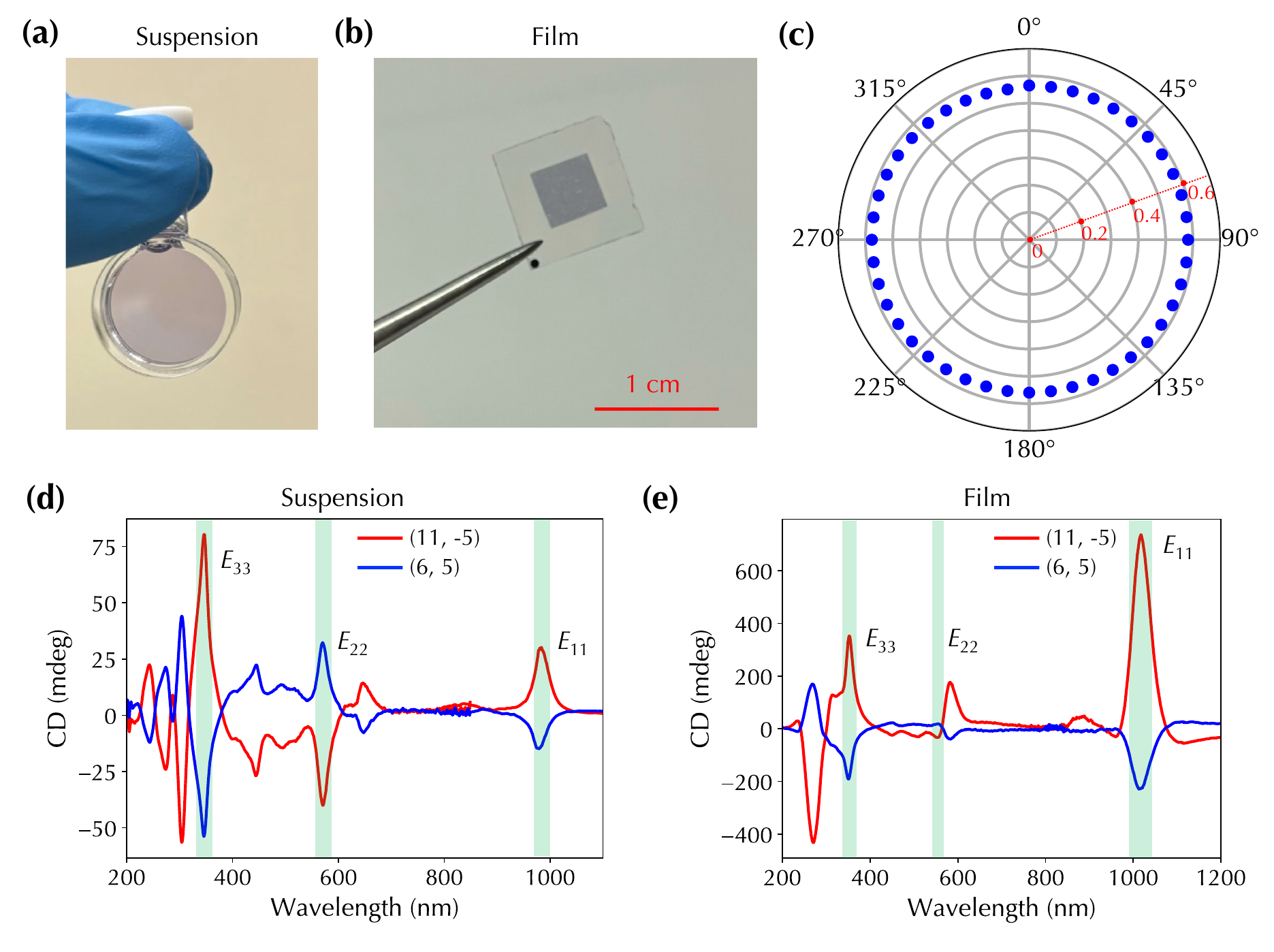}
    \caption{\textbf{Experimental results}. Photos of (a)~separated single-enantiomer suspensions and (b)~wafer-scale films. (c)~Polar plot of polarization-dependent transmission of one obtained film. CD spectra of single-enantiomer (6,5) and (11,-5) (d)~suspensions and (e)~wafer-scale films.}
    \label{fig:exp}
\end{figure}

We verified the optical anisotropy of obtained films by measuring linear-polarization-dependent optical attenuation. We had a laser at a wavelength of 635\,nm (Thorlabs, CPS635R) incident on obtained films. The circular beam size was $\sim2$mm in diameter to probe the degree of alignment in a large area. The linear polarization state of the laser was rotated through a half-waveplate (Thorlabs, WPH10ME-633). The transmitted optical power was measured by a photodiode (Thorlabs, PM16-130). The transmitted optical power of both CNT samples and a bare reference substrate was measured and the ratio was calculated as transmittance. Fig.\,\ref{fig:exp}c shows a polar plot of measured transmittance at various polarization angles and there was no clear angle dependence, confirming that obtained films were randomly oriented. 

\subsection{CD spectra measurement and enantiomeric purity}

We measured the CD and absorption spectra of separated single-enantiomer suspensions and wafer-scale films using a Jasco CD J-1500 spectrometer under normal incidence. The circular probe beam size was $\sim$2\,mm in diameter. The absorption spectra captured the average absorption of LCP and RCP light. The $g$ factor was calculated as the ratio between CD and average absorption. Fig.\,\ref{fig:exp}d shows the measured CD spectra of single-enantiomer (6,5) and (11,-5) suspensions in a wavelength range from 200 to 1200\,nm. The excitonic interband transitions with the same band indices for valance and conduction bands, such as $E_{11}$, $E_{22}$, and $E_{33}$, correspond to the CD spectral features in green shaded rectangles of Fig.\,\ref{fig:exp}d with alternating signs~\cite{SatoEtAl2017PR}.  Other peaks are mainly from cross-polarized excitons~\cite{WeiEtAl2017JACS}. Supplementary Fig.\,2a and 2b show the corresponding absorption and $g$ factor spectra. Based on the ratio of CD and absorption peaks at the $E_{22}$ location, denoted as CD$_\mathrm{norm}$, we estimated enantiomeric purities (EPs) for (6,5) and (11,-5) suspensions following two linear regression fitting curves, EP(6,5) (\%) = $50 + \alpha$CD$_\mathrm{norm}$ and EP(11,-5) (\%) = $50 - \alpha$CD$_\mathrm{norm}$, where $\alpha = 0.421\,\mathrm{mdeg}^{-1}$~\cite{WeiEtAl2017JACS}. Hence, the calculated EPs for (6,5) and (11,-5) suspensions were nearly the same $\sim87\,\%$.

Further, we measured the CD and absorption spectra of wafer-scale single-enantiomer (6,5) and (11,-5) randomly oriented films. In contrast to suspension measurements, the CD measurements of solid-state samples require special treatment to eliminate instrumentation artifacts. We adopted a ``four-configuration'' approach, where CD spectra were captured under four different sample configurations as shown in Supplementary Fig.\,3a, and thus the averaged CD spectra removed artifacts and provided the genuine CD response from samples~\cite{DoumaniEtAl2023NC}; see \emph{Supplementary Information Note 3}, Supplementary Fig.\,3b, and Supplementary Fig.\,3c for more details. Fig.\,\ref{fig:exp}e, Supplementary Fig.\,2c, Supplementary Fig.\,2d display the measured CD, absorption, and $g$ factor spectra of single-enantiomer (6,5) and (11,-5) random films in a wavelength range from 200 to 1200\,nm, respectively. Compared to suspension CD and absorption spectra, as expected, the excitonic peaks in films become broadened due to CNT bundling. Interestingly, the relative heights of $E_{11}$, $E_{22}$, and $E_{33}$ peaks in films are substantially different from those in suspensions, and $E_{11}$ becomes the strongest in films. Moreover, the peak heights of the same excitonic transitions (e.g., $E_{11}$) for two enantiomers are clearly different with a three-times difference, even though the thicknesses of two enantiomer films are nearly the same as evidenced in absorption spectra shown in Supplementary Fig.\,2c and the purities of suspensions are almost identical. We attributed these distinctions in CD spectra between suspensions and films to the formation of local twisted architectures. Although obtained films are randomly oriented when probed in an mm$^2$-size area, local orientations and the twist of orientations in the direction perpendicular to film thickness could occur on a scale of nm$^2$ to $\mu$m$^2$ size during the vacuum filtration process, leading to either negative or positive contributions of CD spectra from twisted stacks of 1D CNT excitonic dipoles~\cite{DoumaniEtAl2023NC}. It is challenging to eliminate the formation of twist structures during film preparation, but we can combine the contributions of CD signals from chiral atomic structures and twisted stacks and treat them as a phenomenological and effective material parameter for designing optoelectronic devices. 

\section{Results and discussion}

\subsection{Transfer matrix and fitting material parameters}

On the macroscopic level, randomly oriented, single-enantiomer CNT films can be treated as a reciprocal, chiral, bi-anisotropic medium with constitutive relations written as $\Vec{D} = \rttensor{\varepsilon}\Vec{E} + \rttensor{\xi}\Vec{H}$ and $\Vec{B} = \rttensor{\zeta}\Vec{E} + \rttensor{\mu}\Vec{H}$, where $\Vec{D}$ is displacement field, $\Vec{E}$ is electric field, $\Vec{H}$ is magnetic field, $\Vec{B}$ is magnetic flux density, $\varepsilon$ is electric permittivity or dielectric function tensor, $\rttensor{\mu}$ is magnetic permeability tensor, and $\rttensor{\xi}$ and $\rttensor{\zeta}$ are two magnetoelectric tensors describing the coupling between electric and magnetic components in constitutive relations~\cite{LindellEtAl1994}. Further, $\rttensor{\xi}$ and $\rttensor{\zeta}$ can be expressed as $\sqrt{\varepsilon_0\mu_0}(\rttensor{\chi} - j\rttensor{\kappa})$ and  $\sqrt{\varepsilon_0\mu_0}(\rttensor{\chi} + j\rttensor{\kappa})$, respectively, where $\rttensor{\chi}$ and $\rttensor{\kappa}$ are two unitless magnetoelectric tensors. Since our produced single-enantiomer CNT films are reciprocal and chiral, $\rttensor{\chi}$ is thus diminishing and $\rttensor{\kappa}$ is specifically called the chirality parameter. 

\begin{figure}
    \centering
    \includegraphics[width=1.0\linewidth]{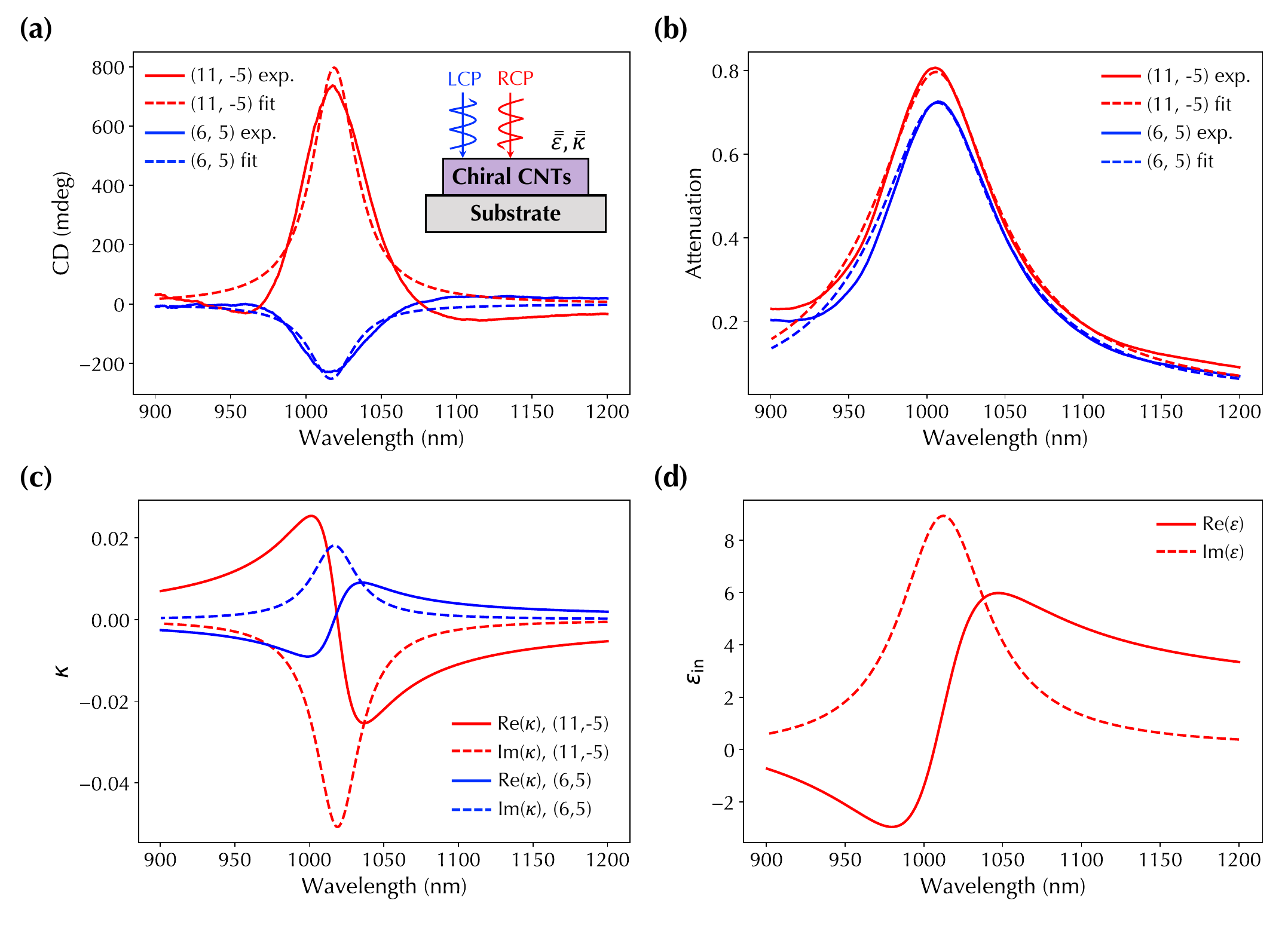}
    \caption{\textbf{Transfer matrix fitting}. Experimental and fitting (a)~CD spectra and (b)~absorption spectra of (6,5) and (11,-5) single-enantiomer films. Extracted complex-valued (c)~chirality parameters and (d)~dielectric functions. }
    \label{fig:fitting}
\end{figure}

In this work, we focused on the $E_{11}$ transitions in (6,5) and (11,-5) films because their CD responses are distant from cross-polarized transitions and have little spectral interference from these transitions. We expressed the relative electric permittivity tensor for a CNT film, $\rttensor{\varepsilon}_\mathrm{CNT}$, as
\begin{align*}
    \rttensor{\varepsilon}_\mathrm{CNT} = 
    \begin{bmatrix}
        \varepsilon_\mathrm{in} & 0 & 0 \\
        0 & \varepsilon_\mathrm{in} & 0 \\
        0 & 0 & \varepsilon_\mathrm{out}
    \end{bmatrix},
\end{align*}
where $\varepsilon_\mathrm{in}$ and $\varepsilon_\mathrm{out}$ are in-plane and out-of-plane dielectric functions. $\varepsilon_\mathrm{in}$ is a function of angular frequency $\omega$ and described using a Lorentz oscillator model as 
\begin{align*}
    \varepsilon_\mathrm{in}(\omega) = \varepsilon_\infty + \frac{\omega_\mathrm{p}^2}{\omega_0^2 - \omega^2 - j\omega\Gamma},
\end{align*}
where $\varepsilon_\infty$ is a constant, $\omega_\mathrm{p}$ is plasma frequency, $\omega_0$ is the resonance frequency for $E_{11}$ transition, and $\Gamma$ is the corresponding decay rate. $\varepsilon_\mathrm{out}$ is a constant as $\sim3.4$ determined in Ref.\,\cite{WuEtAl2024C} since cross-polarization transitions are negligible around the $E_{11}$ transition. In addition, since materials are non-magnetic, $\rttensor{\mu}$ was assumed to be a scalar constant $\mu_0$, which is the vacuum magnetic permeability. 

Moreover, we assumed $\rttensor{\kappa}$ tensor as an isotropic scalar function $\kappa(\omega)$. We would like to note that the CD contribution from chiral atomic arrangements in (6,5) and (11,-5) enantiomers should be isotropic while the CD contribution from twisted stacks could be possibly anisotropic. However, under our current capacity with only normal incidence CD experiments, a close examination of the out-of-plane chirality parameter is not feasible. A carefully crafted custom CD spectrometer with the removal of artifacts and extrinsic effects in solid-state samples will be needed, which is beyond the scope of our current work.

We described $\kappa(\omega)$ using a Condon model~\cite{CondonEtAl1937RMP}. Since $E_{11}$ transitions lie far away from other transitions, the Condon model has one dominant resonance and $\kappa(\omega)$ becomes
\begin{align*}
    \kappa(\omega) = \frac{\omega R}{\omega_{1}^2 - \omega^2 + j\omega\Gamma_{1}},
\end{align*}
where $R$ is the rotational strength of transitions, $\omega_1$ is the resonance frequency for $E_{11}$ transition, and $\Gamma_1$ is the corresponding decay rate. Note that $\omega_1$ is close to but not necessarily identical to $\omega_0$ in dielectric function expression because there is an observable slight difference of the peaks in experimental CD and absorption spectra as shown before. 





We then developed a transfer matrix solver for general bi-anisotropic materials and their layered architectures~\cite{MackayEtAl2022} to fit CD and absorption spectra to extract $\varepsilon_\mathrm{in}(\omega)$ and $\kappa(\omega)$ as illustrated in the inset panel of Fig.\,\ref{fig:fitting}a and Supplementary Fig.\,4. In the medium of the incident light, the electric field vector $\Vec{E}(\Vec{r})$ and magnetic field vector $\Vec{H}(\Vec{r})$ propagating along $z$ axis can be expressed as
\begin{align*}
    \Vec{E}(\Vec{r}) &= \Vec{e}(z)e^{jq(x\mathrm{cos}\psi + y\mathrm{sin}\psi)},\\
    \Vec{H}(\Vec{r}) &= \Vec{h}(z)e^{jq(x\mathrm{cos}\psi + y\mathrm{sin}\psi)},
\end{align*}
where $q = k_0n_1\mathrm{sin}\theta_\mathrm{i}$ is complex-valued wavenumber, $\psi$ is azimuthal angle, $\theta_\mathrm{i}$ is incident angle, $n_1$ is refractive index of incident light medium, $\Vec{e}(z)$ is the unit vector as $\Vec{e_x}(z)\hat{x} + \Vec{e_y}(z)\hat{y} + \Vec{e_z}(z)\hat{z}$, and $\Vec{h}(z)$ is the unit vector as $\Vec{h_x}(z)\hat{x} + \Vec{h_y}(z)\hat{y} + \Vec{h_z}(z)\hat{z}$. For $z$-components, we have $\Vec{e_z}(z) = \nu_{zx}^{ee}\Vec{e_x}(z) + \nu_{zy}^{ee}\Vec{e_x}(z) + \nu_{zx}^{eh}\Vec{h_x}(z) + \nu_{zy}^{eh}\Vec{h_y}(z)$ and $\Vec{h_z}(z) = \nu_{zx}^{he}\Vec{e_x}(z) + \nu_{zy}^{he}\Vec{e_x}(z) + \nu_{zx}^{hh}\Vec{h_x}(z) + \nu_{zy}^{hh}\Vec{h_y}(z)$ because of the coupled electric and magnetic fields in constitutive relations. 

By solving the Maxwell's equations and matching boundary conditions, we have 
\begin{align*}
    \nu_{zx}^{ee} = -\frac{\mu_{zz}\varepsilon_{zx} - \xi_{zz}[\zeta_{zx} + (q/\omega)\mathrm{sin}\psi]}{\varepsilon_{zz}\mu_{zz} - \xi_{zz}\zeta_{zz}},
    \nu_{zy}^{ee} = -\frac{\mu_{zz}\varepsilon_{zy} - \xi_{zz}[\zeta_{zy} - (q/\omega)\mathrm{cos}\psi]}{\varepsilon_{zz}\mu_{zz} - \xi_{zz}\zeta_{zz}}, \\
    \nu_{zx}^{eh} = \frac{\xi_{zz}\mu_{zx} - \mu_{zz}[\xi_{zx} - (q/\omega)\mathrm{sin}\psi]}{\varepsilon_{zz}\mu_{zz} - \xi_{zz}\zeta_{zz}}, 
    \nu_{zy}^{eh} = \frac{\xi_{zz}\mu_{zy} - \mu_{zz}[\xi_{zy} + (q/\omega)\mathrm{cos}\psi]}{\varepsilon_{zz}\mu_{zz} - \xi_{zz}\zeta_{zz}}, \\
    \nu_{zx}^{he} = \frac{\zeta_{zz}\varepsilon_{zx} - \varepsilon_{zz}[\zeta_{zx} + (q/\omega)\mathrm{sin}\psi]}{\varepsilon_{zz}\mu_{zz} - \xi_{zz}\zeta_{zz}}, 
    \nu_{zy}^{he} = \frac{\zeta{zz}\varepsilon_{zy} - \varepsilon_{zz}[\zeta_{zy} - (q/\omega)\mathrm{cos}\psi]}{\varepsilon_{zz}\mu_{zz} - \xi_{zz}\zeta_{zz}}, \\
    \nu_{zx}^{hh} = -\frac{\varepsilon_{zz}\mu_{zx} - \zeta_{zz}[\xi_{zx} - (q/\omega)\mathrm{sin}\psi]}{\varepsilon_{zz}\mu_{zz} - \xi_{zz}\zeta_{zz}}, 
    \nu_{zy}^{hh} = -\frac{\varepsilon_{zz}\mu_{zy} - \zeta_{zz}[\xi_{zy} + (q/\omega)\mathrm{cos}\psi]}{\varepsilon_{zz}\mu_{zz} - \xi_{zz}\zeta_{zz}},
\end{align*}
where $\varepsilon_{ij}, \mu_{ij}, \xi_{ij}, \zeta_{ij}, i,j \in \{x,y,z\}$ are tensor components of $\rttensor{\varepsilon}, \rttensor{\mu}, \rttensor{\xi}$, and $\rttensor{\zeta}$, respectively, for materials in each layer. Furthermore, there are $\mathbf{P_{m}}, \mathbf{M_m}, \mathbf{K_{inc}}$, and $\mathbf{K_{tr}}$ matrices for $m$-th layer with a thickness $d_\mathrm{m}$, which are expressed as
\begin{equation*}
\begin{aligned}
&\mathbf{P_{m}} = \omega
    \left (
    \begin{bmatrix}
        \zeta_{yx} & \zeta_{yy} & \mu_{yx} & \mu_{yy} \\
        \zeta_{yx} & \zeta_{yy} & \mu_{yx} & \mu_{yy} \\
        \zeta_{yx} & \zeta_{yy} & \mu_{yx} & \mu_{yy} \\
        \zeta_{yx} & \zeta_{yy} & \mu_{yx} & \mu_{yy} 
    \end{bmatrix} + 
    \right.
    \\&
    \left.
    \begin{bmatrix}
        \zeta_{yz} + \frac{q}{\omega}\mathrm{cos}\psi & 0 & 0 & 0 \\
        0 & -\zeta_{xz} + \frac{q}{\omega}\mathrm{sin}\psi & 0 & 0 \\
        0 & 0 & -\varepsilon_{yz} & 0 \\
        0 & 0 & 0 & \varepsilon_{xz} 
    \end{bmatrix}
    \cdot \mathbf{I}\cdot 
    \begin{bmatrix}
        \nu_{zx}^{ee} & 0 & 0 & 0 \\
        0 & \nu_{zy}^{ee} & 0 & 0 \\
        0 & 0 & \nu_{zx}^{eh} & 0 \\
        0 & 0 & 0 & \nu_{zy}^{eh} 
    \end{bmatrix} + 
    \right.
    \\&
    \left.
    \begin{bmatrix}
        \mu_{yz} & 0 & 0 & 0 \\
        0 & -\mu_{xz} & 0 & 0 \\
        0 & 0 & -\xi_{yz} + \frac{q}{\omega}\mathrm{cos}\psi & 0 \\
        0 & 0 & 0 & \xi_{xz} + \frac{q}{\omega}\mathrm{sin}\psi & 0
    \end{bmatrix}
    \cdot \mathbf{I}\cdot 
    \begin{bmatrix}
        \nu_{zx}^{he} & 0 & 0 & 0 \\
        0 & \nu_{zy}^{he} & 0 & 0 \\
        0 & 0 & \nu_{zx}^{hh} & 0 \\
        0 & 0 & 0 & \nu_{zy}^{hh} 
    \end{bmatrix} + 
    \right )
\end{aligned}
\end{equation*}
with $\mathbf{I}$ as identity matrix, 
\begin{align*}
    \mathbf{M_m} = e^{j\mathbf{P_m}d_\mathrm{m}},
\end{align*}

\begin{align*}
    \mathbf{K_{inc}} = 
    \begin{bmatrix}
        -\mathrm{sin} \psi & -\mathrm{cos} \psi \mathrm{cos} \theta_\mathrm{i} & -\mathrm{sin} \psi & \mathrm{cos} \psi \mathrm{cos} \theta_\mathrm{i} \\
        \mathrm{cos} \psi & -\mathrm{sin} \psi \mathrm{cos} \theta_\mathrm{i} & \mathrm{cos} \psi & \mathrm{sin} \psi \mathrm{cos} \theta_\mathrm{i} \\
        -\left(\frac{n_1}{\eta_0}\right)\mathrm{cos}\psi \mathrm{cos} \theta_\mathrm{i} & \left(\frac{n_1}{\eta_0}\right)\mathrm{sin}\psi & \left(\frac{n_1}{\eta_0}\right)\mathrm{cos}\psi \mathrm{cos} \theta_\mathrm{i} & \left(\frac{n_1}{\eta_0}\right)\mathrm{sin}\psi \\
        -\left(\frac{n_1}{\eta_0}\right)\mathrm{sin}\psi \mathrm{cos} \theta_\mathrm{i} & -\left(\frac{n_1}{\eta_0}\right)\mathrm{cos}\psi & \left(\frac{n_1}{\eta_0}\right)\mathrm{sin}\psi \mathrm{cos} \theta_\mathrm{i} & -\left(\frac{n_1}{\eta_0}\right)\mathrm{cos}\psi 
    \end{bmatrix} ,
\end{align*}
and
\begin{align*}
    \mathbf{K_{tr}} = 
    \begin{bmatrix}
        -\mathrm{sin} \psi & -\mathrm{cos} \psi \mathrm{cos} \theta_\mathrm{tr} & -\mathrm{sin} \psi & \mathrm{cos} \psi \mathrm{cos} \theta_\mathrm{tr} \\
        \mathrm{cos} \psi & -\mathrm{sin} \psi \mathrm{cos} \theta_\mathrm{tr} & \mathrm{cos} \psi & \mathrm{sin} \psi \mathrm{cos} \theta_\mathrm{tr} \\
        -\left(\frac{n_2}{\eta_0}\right)\mathrm{cos}\psi \mathrm{cos} \theta_\mathrm{tr} & \left(\frac{n_2}{\eta_0}\right)\mathrm{sin}\psi & \left(\frac{n_2}{\eta_0}\right)\mathrm{cos}\psi \mathrm{cos} \theta_\mathrm{tr} & \left(\frac{n_2}{\eta_0}\right)\mathrm{sin}\psi \\
        -\left(\frac{n_2}{\eta_0}\right)\mathrm{sin}\psi \mathrm{cos} \theta_\mathrm{tr} & -\left(\frac{n_2}{\eta_0}\right)\mathrm{cos}\psi & \left(\frac{n_2}{\eta_0}\right)\mathrm{sin}\psi \mathrm{cos} \theta_\mathrm{tr} & -\left(\frac{n_2}{\eta_0}\right)\mathrm{cos}\psi 
    \end{bmatrix}.
\end{align*}

Hence, the overall transfer matrix $\mathbf{Q} = \mathbf{K_{tr}}^{-1}\bigg(\displaystyle\prod_{\mathrm{m}=1}\mathbf{M_m}\bigg)\mathbf{K_{inc}}$ and relates input and output vectors as 
\begin{align*}
    \begin{bmatrix}
        t_\mathrm{s} \\
        t_\mathrm{p} \\
        0 \\
        0 \\
    \end{bmatrix}  = \mathbf{Q}\cdot
    \begin{bmatrix}
        a_\mathrm{s} \\
        a_\mathrm{p} \\
        r_\mathrm{s} \\
        r_\mathrm{p} \\
    \end{bmatrix} = 
    \begin{bmatrix}
        Q_{00} & Q_{01} & Q_{02} & Q_{03} \\
        Q_{10} & Q_{11} & Q_{12} & Q_{13} \\
        Q_{20} & Q_{21} & Q_{22} & Q_{23} \\
        Q_{30} & Q_{31} & Q_{32} & Q_{33} 
    \end{bmatrix} \cdot
    \begin{bmatrix}
        a_\mathrm{s} \\
        a_\mathrm{p} \\
        r_\mathrm{s} \\
        r_\mathrm{p} \\
    \end{bmatrix}.
\end{align*}
Since we have 
\begin{align*}
    \begin{bmatrix}
        r_\mathrm{s} \\
        r_\mathrm{p} \\
    \end{bmatrix}  =
    \begin{bmatrix}
        r_\mathrm{ss} & r_\mathrm{sp} \\
        r_\mathrm{ps} & r_\mathrm{pp}
    \end{bmatrix} \cdot
    \begin{bmatrix}
        a_\mathrm{s} \\
        a_\mathrm{p} \\
    \end{bmatrix},
\end{align*}
and 
\begin{align*}
    \begin{bmatrix}
        t_\mathrm{s} \\
        t_\mathrm{p} \\
    \end{bmatrix}  =
    \begin{bmatrix}
        t_\mathrm{ss} & t_\mathrm{sp} \\
        t_\mathrm{ps} & t_\mathrm{pp}
    \end{bmatrix} \cdot
    \begin{bmatrix}
        a_\mathrm{s} \\
        a_\mathrm{p} \\
    \end{bmatrix},
\end{align*}
we can represent all reflection and transmission coefficients under different combinations of polarization states based on elements of matrix $\mathbf{Q}$. Specifically, we have reflection and transmission coefficients under linear polarizations as
\begin{align*}
    r_\mathrm{ss} = \frac{Q_{30}Q_{23} - Q_{33}Q_{20}}{Q_{22}Q_{23} - Q_{23}Q_{32}}, \quad
    r_\mathrm{sp} = \frac{Q_{31}Q_{23} - Q_{33}Q_{21}}{Q_{22}Q_{23} - Q_{23}Q_{32}}, \\
    r_\mathrm{ps} = \frac{Q_{32}Q_{20} - Q_{30}Q_{22}}{Q_{22}Q_{23} - Q_{23}Q_{32}}, \quad
    r_\mathrm{pp} = \frac{Q_{32}Q_{21} - Q_{31}Q_{22}}{Q_{22}Q_{23} - Q_{23}Q_{32}}, \\
    t_\mathrm{ss} = Q_{00} + Q_{02}r_\mathrm{ss} + Q_{03}r_\mathrm{ps}, \quad
    t_\mathrm{sp} = Q_{01} + Q_{02}r_\mathrm{sp} + Q_{03}r_\mathrm{pp}, \\
    t_\mathrm{ps} = Q_{10} + Q_{12}r_\mathrm{ss} + Q_{13}r_\mathrm{ps}, \quad
    t_\mathrm{pp} = Q_{11} + Q_{12}r_\mathrm{sp} + Q_{13}r_\mathrm{pp}, 
\end{align*}
and coefficients under circular polarizations as
\begin{align*}
    r_\mathrm{LL} = -\frac{r_\mathrm{ss} + r_\mathrm{pp} + j(r_\mathrm{sp} - r_\mathrm{ps})}{2}, \quad
    r_\mathrm{LR} = \frac{r_\mathrm{ss} - r_\mathrm{pp} - j(r_\mathrm{sp} + r_\mathrm{ps})}{2}, \\
    r_\mathrm{RL} = \frac{r_\mathrm{ss} - r_\mathrm{pp} + j(r_\mathrm{sp} + r_\mathrm{ps})}{2}, \quad
    r_\mathrm{RR} = -\frac{r_\mathrm{ss} + r_\mathrm{pp} - j(r_\mathrm{sp} - r_\mathrm{ps})}{2}, \\
    t_\mathrm{LL} = \frac{t_\mathrm{ss} + t_\mathrm{pp} + j(t_\mathrm{sp} - t_\mathrm{ps})}{2}, \quad
    t_\mathrm{LR} = -\frac{t_\mathrm{ss} - t_\mathrm{pp} - j(t_\mathrm{sp} + t_\mathrm{ps})}{2}, \\
    t_\mathrm{RL} = -\frac{t_\mathrm{ss} - t_\mathrm{pp} + j(t_\mathrm{sp} + t_\mathrm{ps})}{2}, \quad
    t_\mathrm{RR} = \frac{t_\mathrm{ss} + t_\mathrm{pp} - j(t_\mathrm{sp} - t_\mathrm{ps})}{2}.
\end{align*}
Hence, we can obtain circular reflectance and transmittance for any polarization combination with $R = |r|^2$ and $T = \frac{n_2\mathrm{Re\{cos\}\theta_\mathrm{tr}}}{n_1\mathrm{cos}\theta_\mathrm{tr}}|t|^2$. Attenuation is defined as $A=-\mathrm{log}_{10}T$, CD is defined as $A_\mathrm{L} - A_\mathrm{R}$, and average absorption is $(A_\mathrm{L} + A_\mathrm{R})/2$. Here, $A_\mathrm{L} = -\mathrm{log}_{10}(T_\mathrm{LL} + T_\mathrm{RL})$ is the attenuation for a LCP incident light and $A_\mathrm{R} = -\mathrm{log}_{10}(T_\mathrm{RR} + T_\mathrm{LR})$ is the attenuation for a RCP incident light. 

We employed the developed transfer matrix to simultaneously fit the CD and absorption spectra of obtained single-enantiomer (6,5) and (11,-5) films and extract material parameters, including $\omega_\infty, \omega_\mathrm{p}, \omega_0, \omega_1, \Gamma, \Gamma_1$, and $R$ in $\varepsilon_\mathrm{in}$ and $\kappa$. We assumed $\varepsilon_\mathrm{in}$ was the same for (6,5) and (11,-5) CNTs while $\kappa$ was different. Fig.\,\ref{fig:fitting}a and \ref{fig:fitting}b display the experimental and fitting CD and absorption spectra for both films, showing a good agreement. Fig.\,\ref{fig:fitting}c and Fig.\,\ref{fig:fitting}d display the extracted complex-valued $\kappa$ and $\varepsilon_\mathrm{in}$, laying the foundation for designing reconfigurable chiral photonic devices. 

\subsection{ML-accelerated calculation and design of reconfigurable chiral photonic devices}

\begin{figure}
    \centering
    \includegraphics[width=1.0\linewidth]{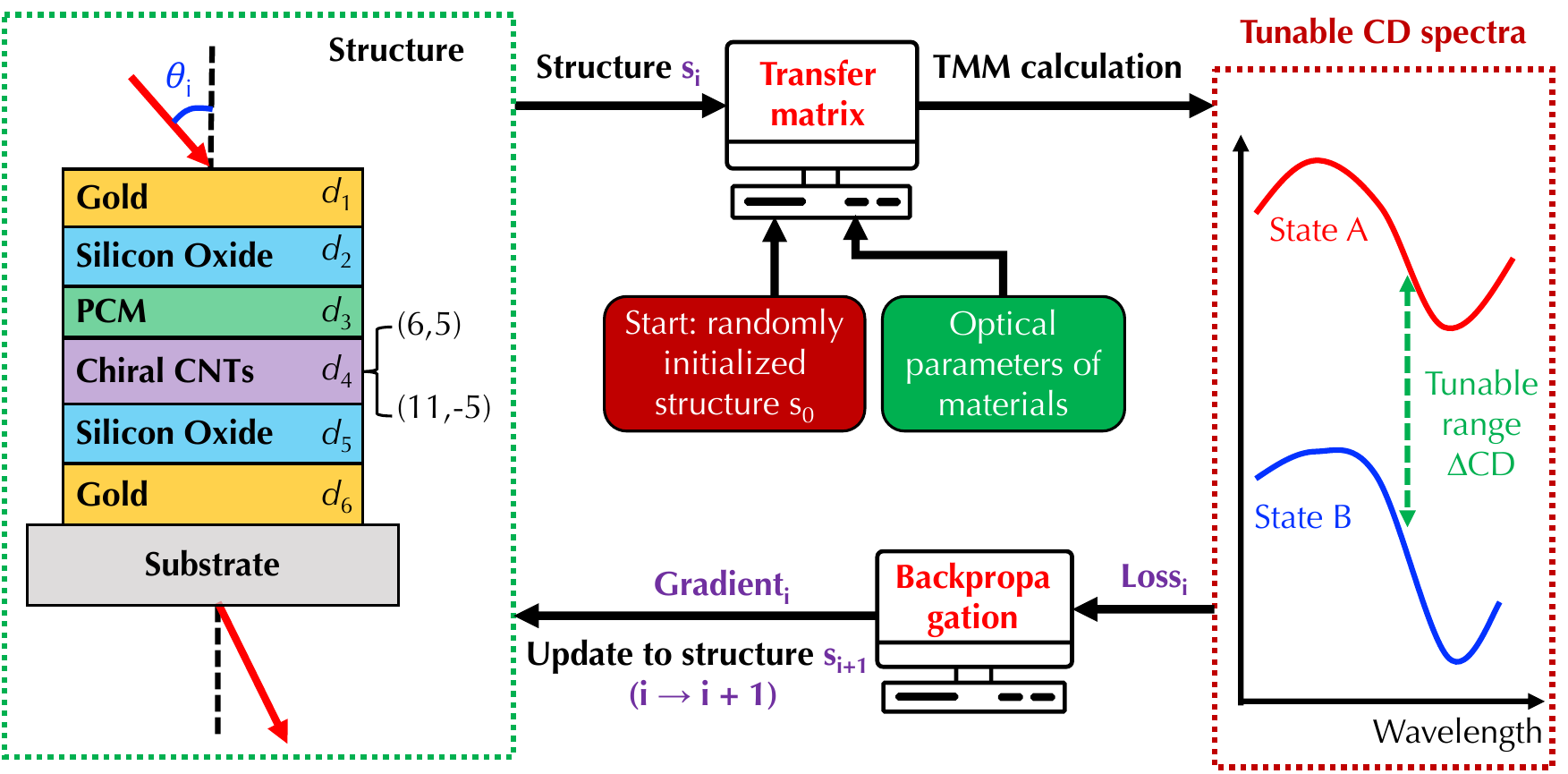}
    \caption{\textbf{Optimization flowchart}. Illustration of employing GPU-accelerated calculations and backpropagation-enabled optimizations for designing reconfigurable chiral photonic devices.}
    \label{fig:opt_diag}
\end{figure}

Since all mathematical operations in transfer matrix calculations are based on tensor algebra, we leveraged the state-of-the-art ML framework, \texttt{PyTorch}, to implement the transfer matrix solver so that the calculations can be substantially accelerated on GPU hardware. Further, we were able to employ the gradient-based backpropagation algorithm in \texttt{PyTorch} to design chiral photonic devices. Fig.\,\ref{fig:opt_diag} illustrates the flowchart of designing a novel heterostack architecture consisting of single-enantiomer (6,5) or (11,-5) CNT randomly oriented films and phase change material (PCM) films for reconfigurable chiral photonic devices. Upon the excitation of electrical or optical pulses, PCMs can be fast and reversely reconfigured between crystalline and amorphous phases, leading to a substantial change of dielectric functions~\cite{WuttigEtAl2017NP,GuoEtAl2019AS,AbdollahramezaniEtAl2020N,ZhangEtAl2021APL}. Furthermore, after the reconfigurability of PCMs, the optical properties of PCMs are preserved even after removing the excitation stimulus. In addition, PCMs are scalable and compatible with other materials for constructing complex architectures~\cite{RaouxEtAl2008JRD}. Such strong non-volatile reconfigurability of scalable PCM films make them ideal for reconfigurable photonic devices in both free space~\cite{WangEtAl2016NP,WangEtAl2021NN,ZhangEtAl2021NN,AbdollahramezaniEtAl2022NC} and integrated circuits~\cite{RiosEtAl2015NP,ChengEtAl2017SA,RiosEtAl2019SA,ZhangEtAl2019NC,FangEtAl2022NN}. The top and bottom gold layers and sandwiched silicon oxide layers are intended to provide multiple reflections to enhance the chiral light-matter interaction. From a practical perspective, silicon oxide layers can also protect PCMs from being exposed to the surrounding environment for a long device lifespan. We chose germanium-antimony-tellurium (GST) PCMs for the demonstration because of its mature manufacturing and broad applications. 

\begin{figure}
    \centering
    \includegraphics[width=1.0\linewidth]{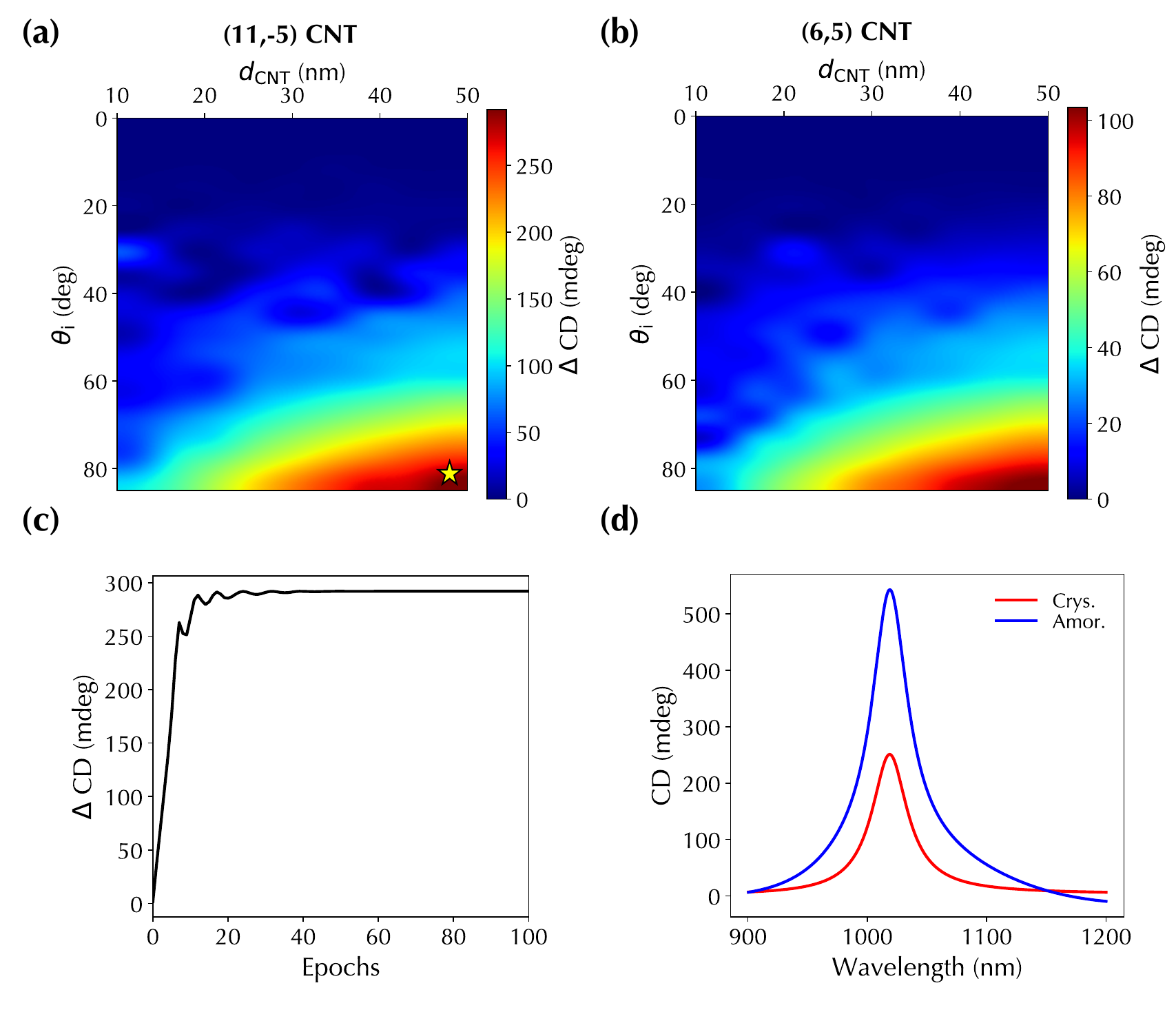}
    \caption{\textbf{Optimization results}. 3D plots of optimal $\Delta$CD after optimization under various $\theta_i$ and CNT thickness $d_\mathrm{CNT}$ for single-enantiomer (a)~(11,-5) and (b)~(6,5) films, respectively. (c)~The optimization curve for the architecture showing the largest tunable at $\theta_i = 85^{\circ}$ and $d_\mathrm{CNT}=50\,$nm. (d)~CD spectra under crystalline and amorphous GST phases for a 50-nm thick (11,-5) film at $\theta_i = 85^{\circ}$ corresponding to the yellow star in (a).}
    \label{fig:opt_result}
\end{figure}

Our design target is to find the best combination of layer thicknesses $\{d_1, d_2, ..., d_6\}$ under a given incidence angle $\theta_\mathrm{i}$ to produce the largest tunable range of CD, denoted as $\Delta$CD. The design started with a randomly initialized structure $s_0$ and known optical constants of materials. Specifically, we utilized the fitted material parameters of single-enantiomer (6,5) and (11,-5) CNT randomly oriented films and dielectric functions of GST, silicon oxide, and gold films from our prior work~\cite{TangEtAl2022LPR} and public database\footnote{Refractive Index Database. https://refractiveindex.info/.}. We performed the forward calculation of CD spectra under crystalline and amorphous GST phases and defined $-\Delta$CD as the loss function. Hence, the minimization of the loss in the backpropagation algorithm is equivalent to maximizing $\Delta$CD. Based on the calculated gradient, the structural parameters were updated to $s_1$ and these cycles were repeated multiple times to achieve optimal design. 


Fig.\,\ref{fig:opt_result}a and \ref{fig:opt_result}b show 3D plots of optimal $\Delta$CD after optimization under various $\theta_i$ and CNT thickness $d_\mathrm{CNT}$ for (11,-5) and (6,5) films, respectively. It is clear that large oblique incident angles and thick films produce strong reconfigurability because oblique incidence breaks mirror symmetry and generates strong extrinsic CD response and thick films naturally have strong chiral light-matter interaction. Notably, there is no reconfigurability of CD response at normal incidence. It seems to contradict the intuition that multiple reflections between gold reflectors can enhance chiral light-matter interaction and changing the dielectric environment between reflectors can lead to CD change. It is noted that reflectors are still planar and symmetric and the incident circularly polarized light changes the handedness, such as from LCP to RCP light or from RCP to LCP light, upon the reflection. Hence, the handedness of circularly polarized light can not be preserved during multiple reflections so there is no enhancement or reconfigurability. 

Fig.\,\ref{fig:opt_result}c shows the evolution of $\Delta$CD with iteration epochs, confirming a successful optimization driven by the backpropagation algorithm. Fig.\,\ref{fig:opt_result}d demonstrates the largest tunable range of CD spectra under crystalline and amorphous GST phases for a 50-nm thick (11,-5) film at $\theta_i = 85^{\circ}$ (yellow star in Fig.\,\ref{fig:opt_result}a), and Supplementary Fig.\,5a and 5b show the reconfigurability of corresponding attenuation and $g$ factor spectra, respectively. The tunable range gradually decreases with decreasing $\theta_i$ as shown in Supplementary Fig.\,6a - 6d.


\section{Conclusions}

In summary, we created an optical material model, developed a high-performance general transfer matrix solver, extracted material parameters, and efficiently designed and optimized a reconfigurable chiral photonic heterostructure with PCMs for wafer-scale single-enantiomer (6,5) and (11,-5) randomly oriented CNT films. The demonstrated integrated hardware and software platform opens new opportunities for exploring both fundamental chiral phenomena and novel applications based on macroscopic solid-state CNT films.

\pagebreak
\noindent\textbf{Acknowledgement} -- J.\,F., B.\,H., H.\,X., R.\,C., and W.\,G. acknowledge the support from the National Science Foundation through Grant Nos.\,2230727, 2235276, 2316627, and 2321366. K.\,Y. acknowledges support from JSPS KAKENHI through Grant Nos. JP23H00259, and JP24H01200, and U.S.-JAPAN PIRE through Grant No. JPJSJRP20221202, Japan, and ASPIRE project through Grant No. JPMJAP2310, Japan.

\bibliographystyle{elsarticle-num} 
\bibliography{weilu}





\end{document}


\begin{frontmatter}



\title{Supplementary Information: Optical modeling, solver, and design of wafer-scale single-enantiomer carbon nanotube films and reconfigurable chiral photonic devices}


\author[inst1]{Jichao Fan}

\affiliation[inst1]{organization={Department of Electrical and Computer Engineering, The University of Utah},
            addressline={50 S Central Campus Drive}, 
            city={Salt Lake City},
            postcode={84112}, 
            state={UT},
            country={USA}}

\author[inst1]{Benjamin Hillam}
\author[inst2]{Cheng Guo}

\affiliation[inst2]{organization={Department of Applied Physics, Stanford University},
            addressline={348 Via Pueblo}, 
            city={Stanford},
            postcode={94305}, 
            state={CA},
            country={USA}}

\author[inst3]{Hiroyuki Fujinami}
\author[inst3]{Shiba Koki}
\author[inst1]{Haoyu Xie}
\author[inst1]{Ruiyang Chen}
\author[inst3]{Kazuhiro Yanagi}
\affiliation[inst3]{organization={Department of Physics, Tokyo Metropolitan University},
            addressline={1-1 Minami-Osawa}, 
            city={Hachioji-shi, Tokyo},
            postcode={192-0397}, 
            country={Japan}}

\author[inst1]{Weilu Gao\corref{cor1}}
\cortext[cor1]{Corresponding author: weilu.gao@utah.edu (Weilu Gao)}

\end{frontmatter}


\section*{Note 1: The preparation of single-enantiomer (6,5) and (11,-5) suspensions}

CoMoCAT CNTs (SG65, Lot \# 704148, MilliporeSigma) were mixed with a 1.0 wt\% aqueous solution of sodium cholate (SC, $\geq$99.0\%, MilliporeSigma) aqueous solution to prepare a 1.25 mg/mL CNT aqueous suspension. The mixture of CNTs and SC was dispersed using a tip ultrasonic homogenizer (Sonifier 250D, Branson) with 30\% output power for three hours at 18\,$^\circ $C. After tip sonication, the mixture was ultracentrifuged at 210,000\,g for two hours using a Hitachi Himac CP80WX ultracentrifuge equipped with a Himac P40ST-2699 rotor to remove metal catalysts, bundled nanotubes, and other impurities. The top 90\% of the supernatant was then collected and mixed with an aqueous solution of sodium dodecyl sulfate (SDS, $>$99.0\%, MilliporeSigma) to prepare a mixed surfactant system containing 2.0\% SDS + 0.5\% SC. 

A two-step enantiomer separation process using different surfactant concentrations was automatically performed at $19 - 20\,^\circ $C using a high-performance liquid chromatography system (NGC chromatography system, BIO-RAD) with columns filled with gel beads (Sephacryl S-200 HR, GE Healthcare). First, the CNT suspension was loaded into the first column equilibrated with 2.0\% SDS + 0.5\% SC solution, and non-adsorbed CNTs containing (6,5) CNTs were collected as filtrate. The filtrate solution was mixed with SC solution to achieve a 0.5\% SDS + 0.5\% SC surfactant solution and used for the subsequent separation process. After loading the CNT suspension into the second column equilibrated with 0.5\% SDS + 0.5\% SC solution, adsorbed CNTs were eluted by using mixed surfactant solution containing sodium lithocholate (LC, $>$98\%, Tokyo Chemical Industry Co., Ltd.), i.e., 0.5\% SDS + 0.5\% SC + $x$\% LC solution, where $x$ is the LC
concentration that was increased stepwise. The high-purity (11,-5) CNTs were collected by adding a gradient solution of 0.5\%SDS + 0.5\%SC + 0.029\% LC solution, and then (6,5) CNTs were collected by adding a gradient solution of 0.5\%SDS + 0.5\%SC + 0.034\% LC solution. 



\section*{Note 2: The preparation of wafer-scale single-enantiomer randomly oriented films}

For film preparation, the surfactants contained in the separated suspension were exchanged to a 0.3\% sodium deoxycholate solution by ultrafiltration (Amicon Ultra-15 with Ultracel-100 membrane, Merck Millipore). A bench-top centrifuge (TOMY LC-200) was used to accelerate the ultrafiltration by rotating the filter tube. This process was repeated multiple times for a complete surfactant exchange. The exchanged suspension was diluted and poured into a 1-inch filtration system equipped with a 50-nm pore-size filter membrane (Alliance Biosystems, Cat\# E0055). The vacuum pump was turned on to create a fast liquid flow, resulting in a randomly oriented CNT film on the membrane surface. Once the polycarbonate filter membrane with deposited CNTs was completely dry, the film was transferred onto a desired target substrate using a wet transfer process. Specifically, a little droplet of water was first dropped onto the target substrate. The CNT film was then inverted, placing the CNT side in touch with the wet substrate while the polycarbonate side remained above. Once the water was gone, the whole sample was placed inside a reflux condenser filled with chloroform liquid vapor to fully and cleanly remove the polycarbonate membrane. The same procedure was repeated by replacing chloroform with acetone to remove any chloroform residues.

\section*{Note 3: CD and absorption spectra measurements}
CD and absorption spectra were measured using a Jasco J-1500 spectrometer, which is equipped with a Xenon light source and a photomultiplier tube detector for the visible range, and a halogen lamp with an InGaAs detector for the near-infrared range. To facilitate four-configuration measurements of solid-state samples, a customized 3D-printed cuvette with an integrated rotor was developed and utilized, allowing in-plane rotation of samples during measurements. Theoretically, the observed circular dichroism (CD) spectra from solid-state samples, denoted as \( \text{CD}_{\text{obs}} \), can be expressed by the equation \( \text{CD}_{\text{obs}} = \text{CD}_{\text{iso}} + 0.5(\text{LD}' \text{LB} - \text{LD} \text{LB}') + \alpha_{\mathrm{c}} (\text{LD}' \sin(2\theta) - \text{LD} \cos(2\theta)) \). The first term, \( \text{CD}_{\text{iso}} \), represents the intrinsic component of CD resulting from chiral excitons in single-enantiomer CNTs. This component is isotropic and does not depend on sample orientation. The second term, \( \text{LD} \text{LB} \), arises from the interference between the linear dichroism (LD) and linear birefringence (LB) responses of the sample. The \( \text{LD} \text{LB} \) term remains unchanged during sample in-plane rotation around an axis perpendicular to the sample plane, but it reverses its sign when the sample is flipped. The third term represents an artifact resulting from the interaction of residual birefringence in the measurement instrument with the sample’s LD and LB. This artifact alters its sign under a \( 90^\circ \) rotation of the sample around the same axis. To isolate \( \text{CD}_{\text{iso}} \), a four-configuration protocol was implemented. Specifically, as illustrated in Supplementary Fig.\,\ref{SI_fig:exp}a, four CD spectra were calculated or measured, labeled as \( \text{CD}_{\mathrm{m1}} \), \( \text{CD}_{\mathrm{m2}} \), \( \text{CD}_{\mathrm{m3}} \), and \( \text{CD}_{\mathrm{m4}} \). The averaged spectra measured during in-plane rotation, expressed as \( 0.5(\text{CD}_{\mathrm{m1}} + \text{CD}_{\mathrm{m2}}) \) or \( 0.5(\text{CD}_{\mathrm{m3}} + \text{CD}_{\mathrm{m4}}) \), effectively eliminate the third term. Therefore, the term \( 0.5(\text{CD}_{\mathrm{m1}} + \text{CD}_{\mathrm{m2}}) \) is equal to \( \text{CD}_{\text{iso}} + 0.5(\text{LD}' \text{LB} - \text{LD} \text{LB}') \), and the term \( 0.5(\text{CD}_{\mathrm{m3}} + \text{CD}_{\mathrm{m4}}) \) is equal to \( \text{CD}_{\text{iso}} - 0.5(\text{LD}' \text{LB} - \text{LD} \text{LB}') \). Hence,  $\text{CD}_{\text{iso}}$ is obtained as $0.25(\text{CD}_{\mathrm{m1}} + \text{CD}_{\mathrm{m2}} + \text{CD}_{\mathrm{m3}} + \text{CD}_{\mathrm{m4}}).$

\newpage
\begin{figure}
    \centering
    \includegraphics[width=1.0\linewidth]{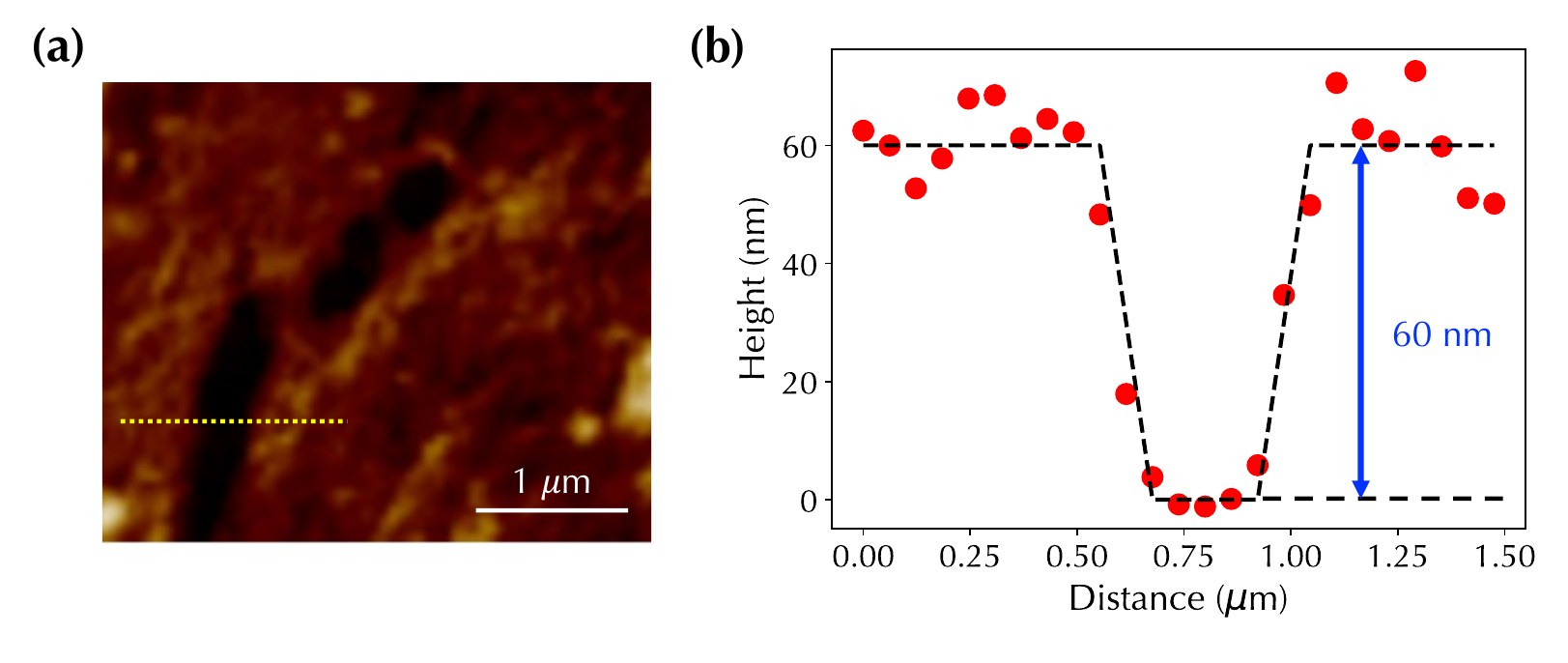}
    \caption{(a)~AFM image of a representative film and (b)~height profile across the yellow dashed line in (a).}
    \label{SI_fig:afm}
\end{figure}

\begin{figure}
    \centering
    \includegraphics[width=1.0\linewidth]{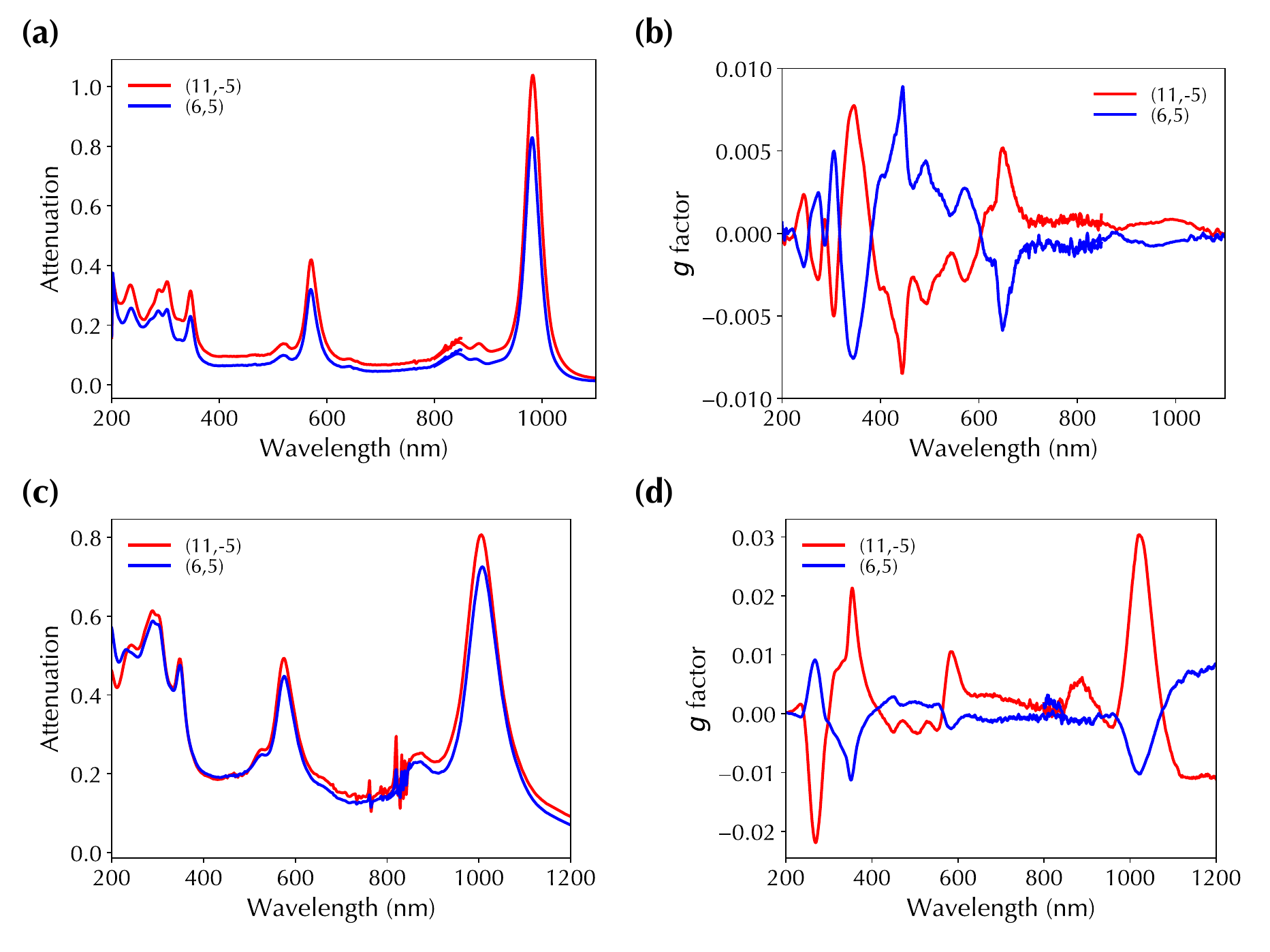}
    \caption{Additional experimental data of (a)~absorption and (b)~$g$ factor spectra for single-enantiomer (6,5) and (11,-5) suspensions. (c) and (d) are corresponding spectra for films.}
    \label{SI_fig:exp}
\end{figure}

\begin{figure}
    \centering
    \includegraphics[width=1.0\linewidth]{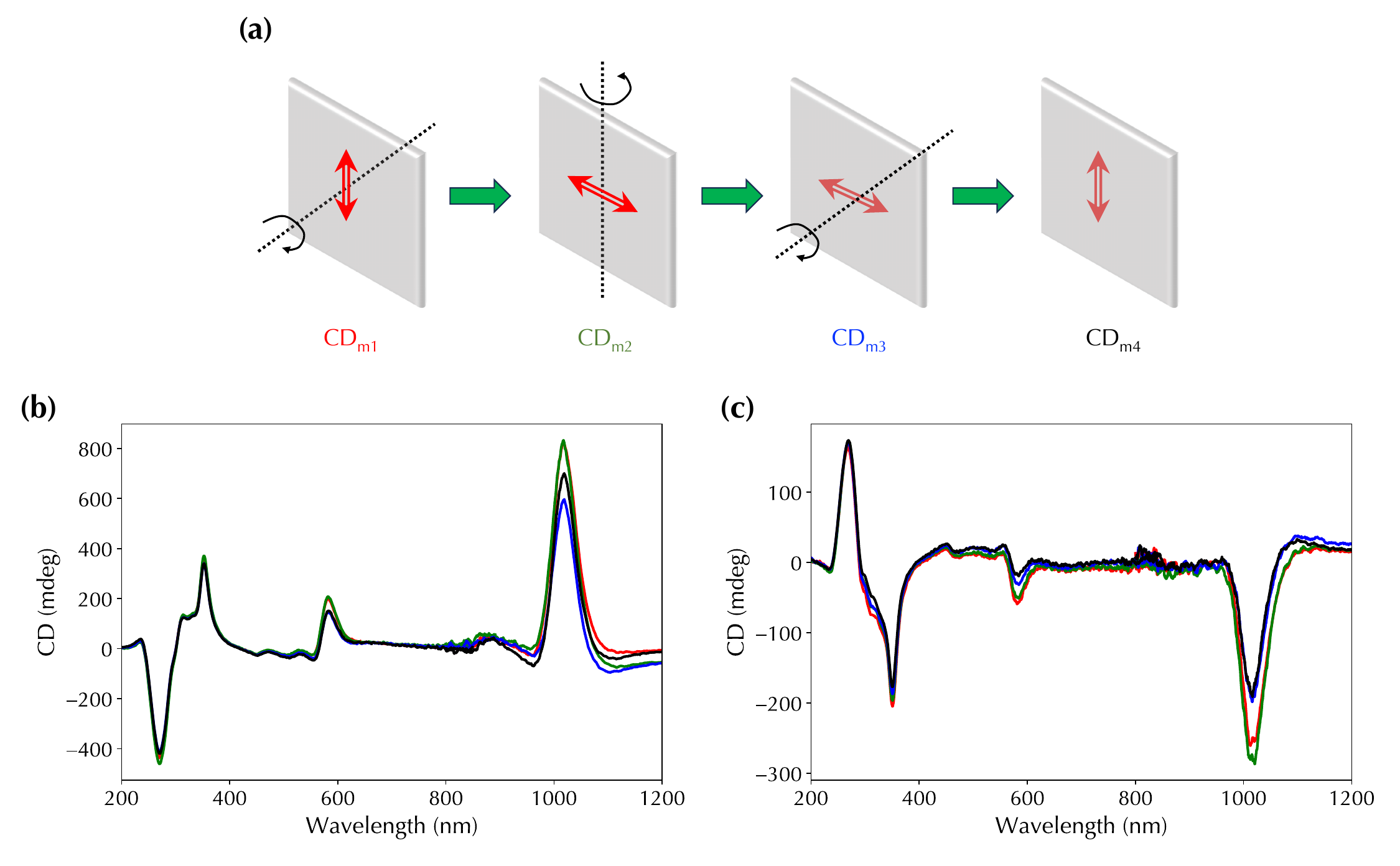}
    \caption{(a)~Illustration of four-configuration CD measurements. Measured CD spectra under four sample configurations for (b)~(11,-5) and (c)~(6,5) films.}
    \label{SI_fig:exp}
\end{figure}

\begin{figure}
    \centering
    \includegraphics[width=0.5\linewidth]{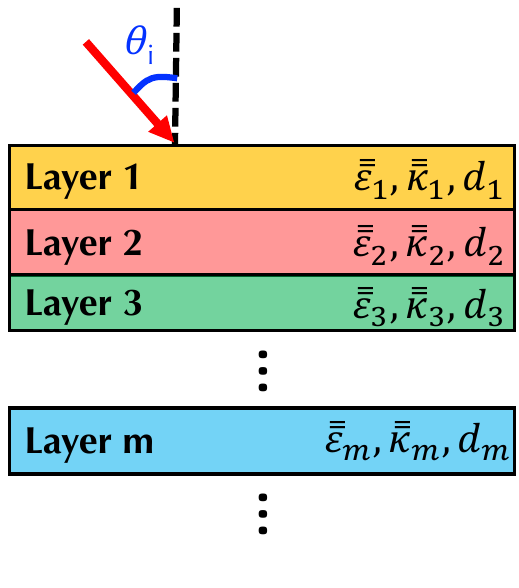}
    \caption{Illustration of a general transfer matrix for layered structures with bi-anisotropic materials.}
    \label{SI_fig:trans_mat}
\end{figure}

\begin{figure}
    \centering
    \includegraphics[width=1.0\linewidth]{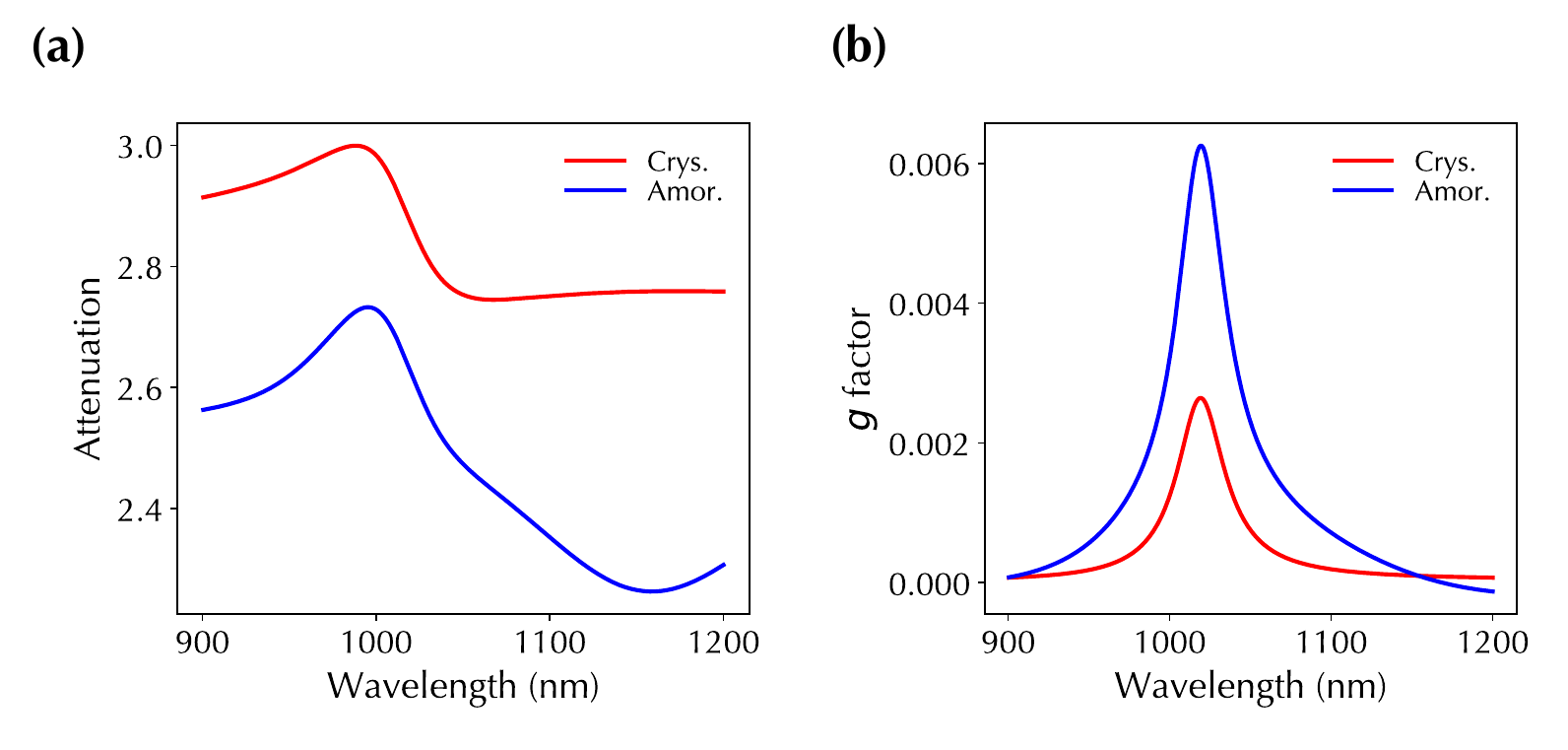}
    \caption{(a)~Absorption and (b)~$g$ factor spectra under crystalline and amorphous PCMs for a layered architecture with $\theta_i = 85^{\circ}$ and 50-nm thick CNT films.}
    \label{SI_fig:opt_best}
\end{figure}

\begin{figure}
    \centering
    \includegraphics[width=1.0\linewidth]{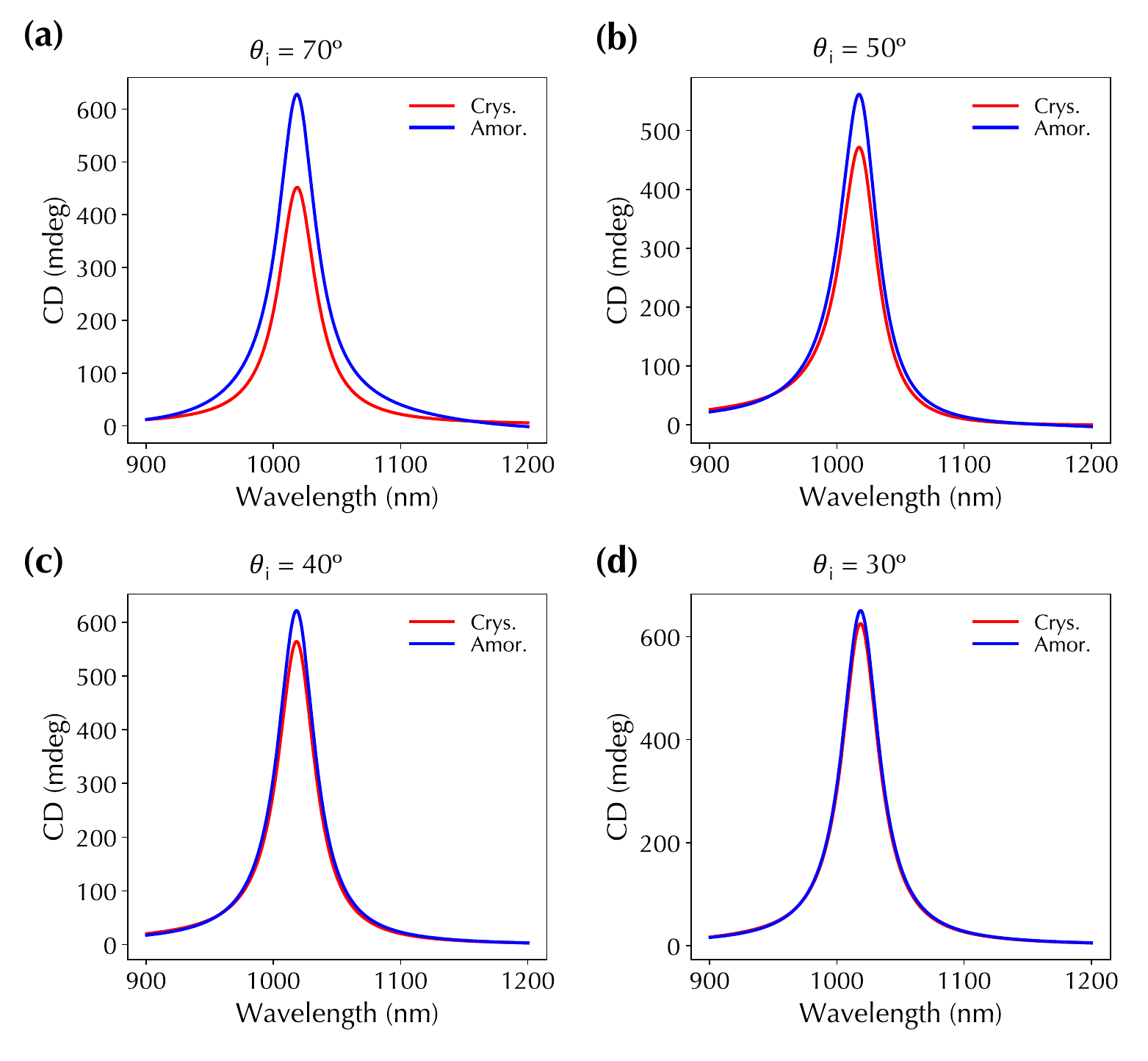}
    \caption{CD spectra under crystalline and amorphous PCMs for a layered architecture with 50-nm thick CNT films for (a)~$\theta_i = 70^{\circ}$, (a)~$\theta_i = 50^{\circ}$, (a)~$\theta_i = 40^{\circ}$, and (a)~$\theta_i = 30^{\circ}$, respectively.}
    \label{SI_fig:opt_angle}
\end{figure}





